\newif\ifTwoColumn%
\newif\ifSUBMIT%
\newif\ifCOMMENTS%
\newif\ifFIGs%
\newif\ifFIGoneColumn%
\let\ifSUBMIT\iftrue%
\let\ifCOMMENTS\iffalse
\let\ifFIGoneColumn\iftrue%
\documentclass[journal=jpclcd,manuscript=letter]{achemso}
\setkeys{acs}{articletitle = true}
\setkeys{acs}{doi = true}
\mciteErrorOnUnknownfalse
\setkeys{acs}{etalmode = truncate, maxauthors = 1000}

\usepackage{graphicx}
\usepackage{dcolumn}
\usepackage{bm}
\usepackage{amssymb}   
\usepackage{amsmath}
\usepackage{xcolor}
\usepackage{todonotes}
\usepackage{menukeys}
\usepackage{hyperref}
\usepackage{xr}
\usepackage{siunitx}
\usepackage{acronym}
\usepackage{array,mathtools,amssymb,booktabs}
\usepackage{rotating}
\usepackage{physics}
\usepackage{acronym}
\usepackage{placeins}

\ifSUBMIT%
  \ifCOMMENTS%
  \else

  \fi
\else
\fi

\title[PNNPs are PNNs]%
{Polymer-Linked Nanoparticle Networks Running on Heat
Can Act as Computing Devices}

\author{Xingfei Wei}
\affiliation{Department of Chemistry, Johns Hopkins University, 
Baltimore, Maryland 21218, USA}

\author{Manuel Palma Banos}
\affiliation{Department of Chemistry, Johns Hopkins University, 
Baltimore, Maryland 21218, USA}

\author{Rigoberto Hernandez}
\email{r.hernandez@jhu.edu}
\affiliation{Department of Chemistry, Johns Hopkins University, 
 Baltimore, Maryland 21218, USA}
\alsoaffiliation{Department of Chemical \& Biomolecular Engineering, 
Johns Hopkins University, Baltimore, Maryland 21218, USA}
\alsoaffiliation{Department of Materials Science \& Engineering, 
Johns Hopkins University, Baltimore, Maryland 21218, USA}


\date{\today}

\begin{document}
\newlength\figurewide
\ifFIGoneColumn
  \figurewide=.5\columnwidth
\else
  \figurewide=.9\columnwidth
\fi

\begin{tocentry} 
\ \\
  \includegraphics[clip=true,clip=true,height=5cm]{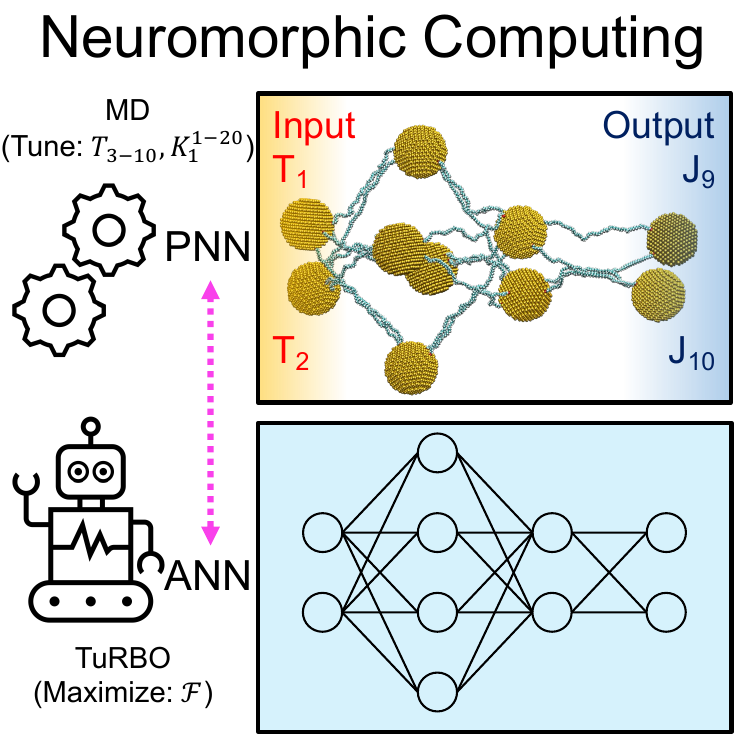}
\ \\
\end{tocentry}

\begin{abstract} 

Developing physical neural network (PNN) hardwares is important to
next generation artificial intelligence systems.
Phononic devices---using heat current to encode
and process information---is one of the solutions to neuromorphic computing.
In this work,
we back map an artificial neural network (ANN)
into a PNN simulation model using polymer networked nanoparticles (PNNPs).
Our atomistic simulation results demonstrate that the polymer linked nanoparticle networks
can potentially realize information processing using heat current.
Using high-throughput molecular dynamics (MD) simulations
and the trust region Bayesian optimization (TuRBO) methods,
we tune the plasticity of polymer linkers
and the temperatures of nanoparticles
to optimize the performance of the PNNP machines,
which is similar to tune the weights and bias in ANNs.
After 5 rounds of high-throughput MD simulations,
we show that the PNNP machines have improved in performance.
We also use a testing data set to verify
the heat flow outputs from the top 5 PNNP machines in each round.

\end{abstract}

\maketitle


\acrodef{ACCESS}{Advanced Cyberinfrastructure Coordination Ecosystem: Services \& Support}
\acrodef{AuNP}{gold nanoparticle}
\acrodef{FDR}{Fisher's Discriminant Ratio }
\acrodef{ANN}{artificial neural network}
\acrodef{PNN}{physical neural network}
\acrodef{PNNP}{polymer networked nanoparticles}
\acrodef{CNN}{convolutional neural network}
\acrodef{NP}{nanoparticle}
\acrodef{ENP}{engineered nanoparticle}
\acrodef{MD}{molecular dynamics}
\acrodef{ACM}{Autonomous Computing Material}
\acrodef{AA}{All-Atom}
\acrodef{MC}{Monte Carlo}
\acrodef{ML}{Machine Learning}
\acrodef{CG}{Coarse-Grained}
\acrodef{LAMMPS}{Large-scale Atomic Molecular Massively Parallel Simulator}
\acrodef{OPLS}{Optimized Potentials for Liquid Simulations}
\acrodef{LJ}{Lennard-Jones}
\acrodef{SI}{Supplementary Information}
\acrodef{MFT}{Mean Field Theory}
\acrodef{GPU}{graphics processing unit}
\acrodef{AI}{artificial intelligence}
\acrodef{BO}{Bayesian optimization}
\acrodef{TuRBO}{trust region BO}

The first \ac{ANN} model
was developed in 1943 by McCulloch and Pitts,
to represent the activity in neuron 
systems.\cite{mcculloch43,datamapu23,fleury25}
In 1949, Donald Hebb proposed
the Hebbian learning mechanism
demonstrating a possible realization for the
tuning of weights between neurons in \acp{ANN}.\cite{hebb49,fleury25}
Since then, neural network models have been applied 
broadly.\cite{datamapu23}
For example,
the \ac{CNN} architecture has proven to be
highly successful in image processing.\cite{khan20}
Meanwhile, the computational cost for training neural networks 
increases significantly with the increasingly larger models
needed for modern applications.
The first \ac{CNN} model---viz LeNet---%
developed by \citet{lecun89} in 1989,
was trained on an IBM 486 PC with
a Intel 80486 microprocessor and an AT\&T digital signal processor card.
In 2012, the well-known AlexNet---a deep \ac{CNN}---%
was trained on 2 NVIDIA GeForce GTX 580 graphic cards
with 3 GB of RAM on each.\cite{alexnet12}
The GPT-3 model published in 2020 was trained on $\sim$10,000 NVIDIA V100 \acp{GPU}
with 32 GB of RAM on each.\cite{brown20gpt3}
Currently, \ac{AI} systems are built on deep \acp{ANN}
using a significant amount of \acp{GPU} resources.
However,
crucial challenges---%
viz. the von Neumann bottleneck, limited data storage capacity,
and high energy consumption---%
are limiting traditional computers.\cite{Williams2017,Shalf2020,Sebastian2020,Pronold2022,fleury25}

A possible resolution to the imminent bottlenecks from
existing computer architectures comes from the realization of
networks of unconventional computing materials.\cite{fleury25}
Various electronic and optical devices have been developed
that exhibit the
requisite computing capabilities---that is,
performing computation, providing data storage
of the capability to train \ac{AI} models---%
in \acp{PNN}.\cite{Sebastian2020,wetzstein20,fleury25}
The first \ac{PNN} device---the Mark I Perceptron---was 
developed by Rosenblatt in 1958.
It is an electronic device and
can recognize simple digital numbers.\cite{rosenblatt57,rosenblatt58}
Recently, memristors have been used to realize in-memory computing
while potentially being more energy efficient
and perhaps breaking the 
von Neumann bottleneck.\cite{williams2008,Sebastian2020,Xiao2023,Weilenmann2024}
Developing \acp{PNN}
using optical components represents another emerging
area offering the possibility of 
faster speeds and 
increased energy efficiency.\cite{wetzstein20,mcmahon23a,Li2023mem}
For example,
Wang et al.~reported a multilayer optical \ac{PNN}
capable of performing
image classification and object detection.\cite{mcmahon23b}

Alternatively, \acp{PNN} could be realized with nanoscale
components that can effectively act as \acp{ACM}.\cite{hern21b}
Helou et al.\cite{ElHelou2021}
developed soft electrically conductive polymer networks
that can act as digital logic gates.
Our group has shown that \acp{PNNP} are 
\acp{ACM}
offering the primitive properties needed
for data storage and computing 
applications.\cite{hern21i,hern23e,hern24k,hern25i}
Using theoretical and simulation models,
such \acp{PNNP} under
applied electrical fields were seen
to be capable of the regulation of the network connections in
2D and 3D regular arrays of polymer electrolytes linked nanoparticles
needed for \ac{ACM} applications.\cite{hern21d,hern23e,hern24h}
The use of 
heat current in phononic devices
to realize logic gate and information processing
while controlling thermal energy management
has also been realized in models by other
groups.\cite{bli2012,bli2020, bli2021,bli2022}
Using \ac{MD} simulations, 
we have further demonstrated that
different types of \acp{PNNP} can regulate
heat transfer.\cite{hern22l}
Examples of these materials included
tree-structured polymer networks\cite{hern23i}
and electrically gated ferrocenyl molecules.\cite{hern24j,hern25j}

\begin{figure}
\includegraphics[clip=true,width=0.8\linewidth]{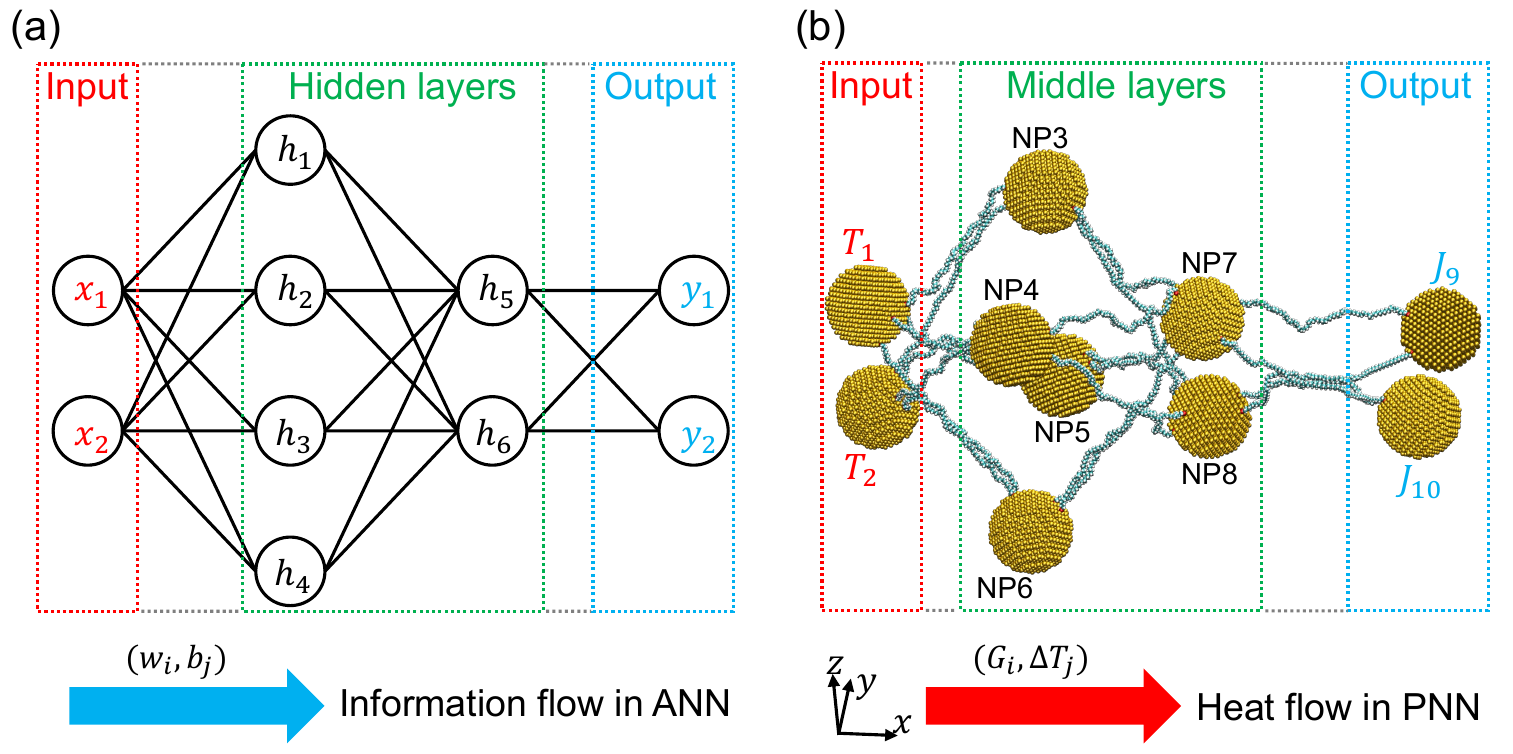}
\caption{
(a) The scheme of a simple 2-4-2-2 \ac{ANN} network 
used here 
to classify (or label) the
input data ($x_1$, $x_2$) according to the output data ($y_1$, $y_2$).
(b) The analogous \ac{PNNP} model of 20 polymers linking 10 \acp{AuNP} for
classifying input temperatures ($T_1$, $T_2$)
according to the output heat currents ($J_9$, $J_{10}$).
}
\label{fig:model}
\end{figure}

In this work, we develop a high-throughput
all-atom \ac{MD} simulation workflow
to demonstrate that \acp{PNNP}---consisting of
polymer linked \acp{AuNP},---%
relying on heat flow
can potentially perform information processing.
The \ac{PNNP} models are illustrated in Figure~\ref{fig:model}b,
and also available in 
Figures~S1 and ~S2 in the \ac{SI}.
The \ac{PNNP}  physical parameters are
temperatures (biases) of the \acp{AuNP}
and plasticities (weights) of the polymer linkers.
The structure is inspired and supported by the corresponding
\acp{ANN} shown in Figure~\ref{fig:model}.

We apply the 
\ac{BO} algorithm known as the
\ac{TuRBO} method\cite{eriksson19turbo}
to optimize the physical parameters and train \acp{PNNP}
in 5 rounds of high-throughput \ac{MD} simulations.
Specifically,
we start the first round with 400 randomly selected parameter settings 
in warm start conditions,
and followed by 4 rounds of the \ac{TuRBO} optimization
400 machines.
We find that the performance of 
the \acp{PNNP} has been improved,
using this workflow,
as one would expect from
the analogous training of the \acp{ANN}.
We also validate the performance of the top 5 machines in each round,
using a testing data set of 100 new inputs.

{\it Training PNNP machines.}
We aim to train \acp{ANN} and \acp{PNN} that can map different temperature input
settings  ($T_1$, $T_2$)
to targeted heat current outputs ($J_9$, $J_{10}$),
by varying the internal temperatures and the connection weights,
as indicated in Figure~\ref{fig:model}, and described in detail below.
For illustration, we constructed 
analogous \acp{ANN} and \acp{PNNP} machines which 
reduce the assignment of temperature pairs
to be inside or outside a certain circle of temperatures
from a quadratic function to a linear one.
Specifically, 
we constructed a database of 200 randomized auxiliary points ($x_1$,$x_2$)
for which 100 points have radius less than 0.5 
(and a categorical label equal to 0 or inside)
and 100 points have radius 0.7 to 1.0
(and a categorical label equal to 1 or outside.)
Details governing the probability distributions for the selection
of these points may be found in the \ac{SI}.
The corresponding temperature pairs in 
the \ac{ANN} training data are then obtained from 
($x_1$,$x_2$)
using $T=25x+300$,
and shown in Figure~S3 in the \ac{SI}.
In the \acp{ANN} of Figure~\ref{fig:model}a,
the prediction of the categorical value
from the output node ($y_1$,$y_2$) is taken 
as 0 if $y_1 < y_2$, and 1 otherwise.
For the \acp{PNNP} of Figure~\ref{fig:model}b,
the training is similar to training of the\acp{ANN},
except the input data is ($T_1$, $T_2$) 
the output data is ($J_9$, $J_{10}$), and we
employ a slightly different label for the supervised training.
In principle, our \ac{PNNP} machines could also be trained to achieve
a given input ($T_1$, $T_2$) to be
labeled 0 if $J_9 <J_{10}$ and 1 otherwise,
corresponding to the label for the \acp{ANN}.
For simplicity,
we used a weaker constraint by training on 
the absolute difference $|J_9 - J_{10}|$
rather than on the specific outputs ($J_9$, $J_{10}$).
Larger outputs of $|J_9 - J_{10}|$ values are assigned
to an auxiliary group $A$ 
and the smaller ones to group $\overline{A}$.
The \ac{FDR}, ${\cal F}$, is calculated 
as a measure of the difference between groups $A$ and $\overline{A}$,
and the training of the \ac{PNNP} process 
maximizes ${\cal F}$ without specifying whether group 
$A$ is 0 (inside) or 1 (outside) as long as it correctly sorts the points
into the two groups.
Once sorted, we can label $A$ to be 0 or 1 according to whether it 
represents inside or outside points, respectively.
We randomly selected 6 input data entries
(3 inside and 3 outside) from the \ac{ANN} training data set to train \acp{PNNP},
and name them as inputs 1-6 in Table~\ref{tab:input_values};
see Figure~S3 in the \ac{SI}.
The \ac{PNNP} training data set is much smaller because high-throughput \ac{MD} simulations 
are computationally expensive,
and we restricted the size accordingly with seemingly
little loss of generality as reported below.

\begin{figure}
\includegraphics[clip=true,width=0.7\linewidth]{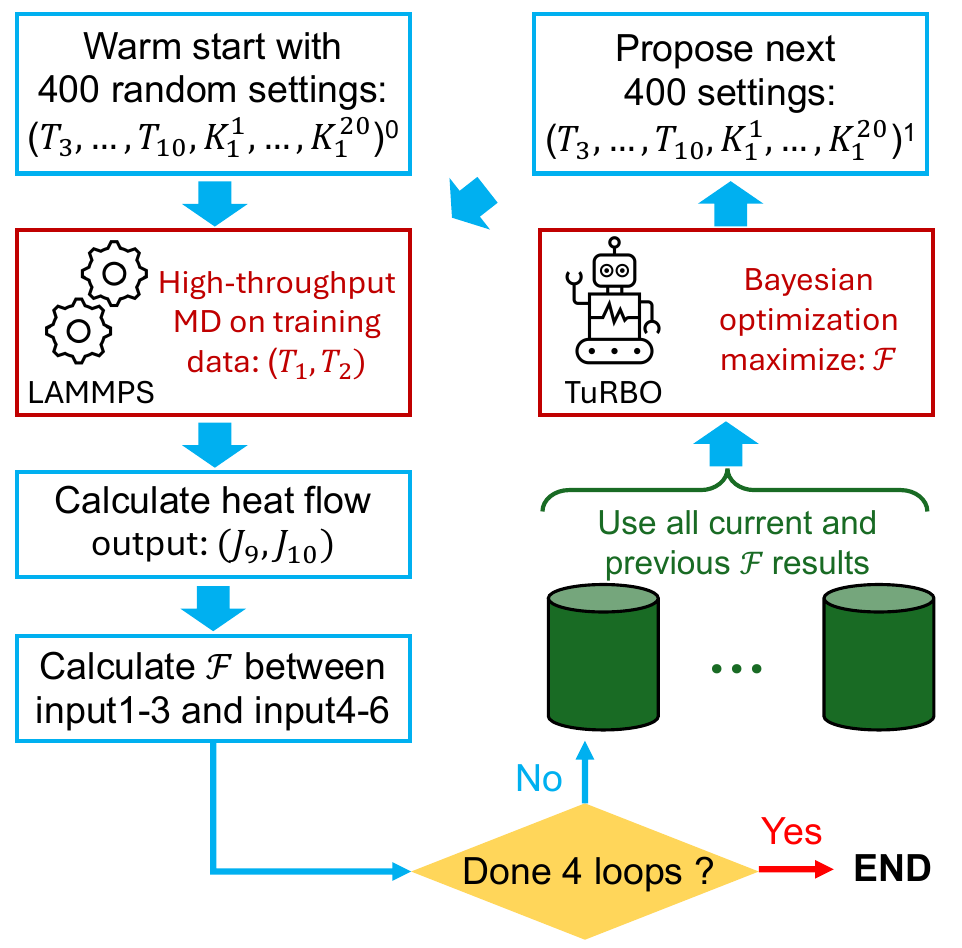}
\caption{
Scheme for training \acp{PNNP} using \acp{AuNP},
where ($T_{3}$, ..., $T_{10}$, $K_{1}^{1}$, ..., $K_{1}^{20}$)$^0$
means the initial warm start temperature and dihedral energy settings.
The process starts with a warm start of 400 randomly selected settings.
Thereafter, 4 loops using \ac{TuRBO} Bayesian optimization and
high-throughput MD simulations were performed.
}
\label{fig:scheme}
\end{figure}

\begin{table}[htbp]
\centering
\caption{The temperature settings for input 1-6. Here, $x_1$ and $x_2$ are ANN model inputs,
$y_1$ and $y_2$ are the ANN model outputs raw score `logits',
and $T_1$ and $T_2$ are the PNNP model temperature inputs. }
\label{tab:input_values}
\begin{tabular}{cccS[table-format=3.2]S[table-format=3.2]cccc}
\toprule
Input & {$x_1$} & {$x_2$} & {$T_1$ (K)} & {$T_2$ (K)} &Label & Category\\
\midrule
1 & -0.1997 & -0.1384 & 295.0 & 296.5 & 0 & inside \\
2 & -0.0617 & 0.0462 & 298.5 & 301.2 & 0 & inside \\
3 & 0.0843 & 0.1937 & 302.1 & 304.8 & 0 & inside \\
4 & 0.7473 & -0.3677 & 318.7 & 290.8 & 1 & outside \\
5 & 0.7387 & 0.3338 & 318.5 & 308.3 & 1 & outside \\
6 & -0.5639 & -0.4374 & 285.9 & 289.1 & 1 & outside \\
\bottomrule
\end{tabular}
\end{table}

For simplicity, we define a reference machine
using a \ac{PNNP}
whose settings correspond to
those for polyethylene linkers
and no temperature bias is applied
to the gaps across $T_3$ to $T_{10}$.
Its the resulting accumulative thermal energy 
measured at NP9 and NP10 is
shown in Figure~S4.
%
%
The slopes of these curves correspond to the heat current ($J_9$, $J_{10}$).
However, Figure~S5a shows that
after 10 independent simulations,
the ($J_9$, $J_{10}$) outputs are all similar
across six selected inputs for the reference \ac{PNNP} machine.
Consequently, the \ac{PNNP} requires training to discern between the
inputs, and that is reported below.


We calculated $|J_9-J_{10}|$ and \ac{FDR}, ${\cal F} $, shown in Figure~S5b,
to provide a figure of merit useful for training.
The \ac{FDR} value, ${\cal F}$, in Eq.~\ref{eq:fdr} is an easy way to
quantitatively measure the difference between
2 groups of $|J_9-J_{10}|$ outputs and
is the target to optimize the \ac{PNNP} performance.
Specifically, the \ac{FDR} is
\begin{equation}
{\cal F} = \frac{(\mu_a - \mu_b)^2}{\sigma_a^2 + \sigma_b^2}, 
\label{eq:fdr}
\end{equation}
where $\mu_a$ and $\mu_b$ are the average of
all $|J_9-J_{10}|$ data from inputs 1-3 and inputs 4-6, respectively,
and $\sigma_a$ and $\sigma_b$ are the corresponding standard deviations.
A small ${\cal F} $ value of 0.0083 and the similar $|J_9-J_{10}|$ distributions in Figure~S5b
indicate that the reference \ac{PNNP} machine using polyethylene linkers can not
realize computing.

Figure~\ref{fig:scheme} shows the work flow of using the \ac{TuRBO} algorithm
to optimize the \ac{PNNP} models by
tuning the plasticities/weights of the 20 polymers ($K_1^{1}$, ..., $K_1^{20}$)
and changing the temperature settings of the 8 \acp{AuNP} ($T_3$, ..., $T_{10}$),
28 parameters in total.
The physical rational for training \ac{PNNP} using heat current for information processing
is based on the heat conductance expression,\cite{hern22l,hern23i}
\begin{equation}
J = G\times \Delta T,
\label{eq:heat}
\end{equation}
where $J$ is the heat flow rate,
$G$ is the heat conductance across 2 \acp{AuNP},
and $\Delta T$ is the temperature gap.
In general, the phonon heat current though a molecular chain at temperature $T$
can be calculated by the Landauer formula,\cite{markussen09,nitzan2023b,ztian26}
\begin{equation}
J(T) = \frac{\hbar}{2\pi} \int_{0}^{\infty} d\omega~\omega \Xi(\omega) [n(T_i) - n(T_j)],
\label{eq:landauer}
\end{equation}
where $\omega$ is the phonon frequency, $\Xi(\omega)$ is the phonon transmission,
and $n = \left[ \exp\left( \frac{\hbar \omega}{k_{\text B} T} \right) - 1 \right]^{-1}$
is the Bose-Einstein distribution of phonons
on the 2 linked \acp{AuNP} with index $i$ and $j$.
Zimbovskaya and Nitzan used the non-equilibrium Green's function
to solve $\Xi(\omega)$ and calculate heat conductance through a polymer chain.\cite{nitzan2023b}
In our \ac{MD} simulation models,
we tune the polymer plasticity (dihedral torsion energies) 
and temperatures on \acp{AuNP}
to change the heat conductance $G$ and heat current;
see Figures~S6 -~S8 in the \ac{SI}.
Although stretching polymers\cite{jliu2012,tluo2012,nitzan2020}
or using more polymer linkers\cite{hern22l,nitzan2022}
can significantly increase the heat current, in this work,
we do not explore these additional parameters in training the \ac{PNNP} model.

\begin{figure}
\includegraphics[clip=true,scale=1,width=0.7\linewidth]{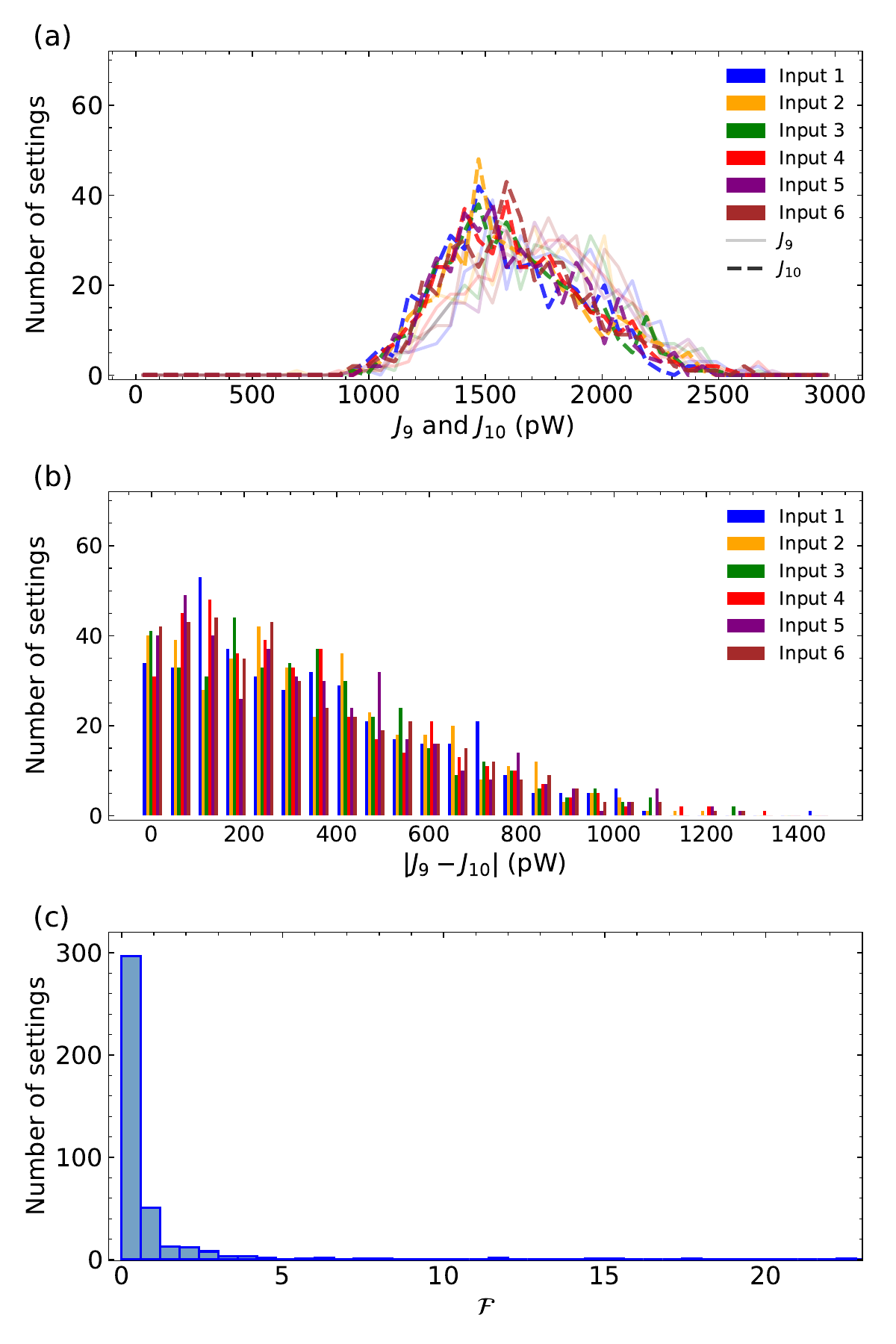}
\caption{
Output results of the PNNP machines at the end of the first round warm start
of \ac{MD} simulations in the scheme shown 
in Figure \ref{fig:scheme}:
(a) Heat current flow rate distributions, $J_9$ and $J_{10}$, at NP9 and NP10 for the 6 different inputs.
(b) Frequency distributions for the absolute
heat current differences, $|J_9-J_{10}|$,
for 6 different input values.
Each set of 6 bars correspond to the same histogram bin width
equals to 60 pW.
(c) Frequency distribution for the
\ac{FDR}, $\cal F$, defined in Eq.~\ref{eq:fdr},
over the ensemble of 400 parameter settings.
}
\label{fig:round1_fdr}
\end{figure}

{\it Comparing outputs from PNNP machines.}
We first generate a perturbed ensemble of 400 \ac{PNNP} machines
by randomly perturbing parameters from a 
configuration of the reference \ac{PNNP} machine 
to determine whether or not such an ensemble 
contains only machines that like the reference cannot 
process information.
Specifically, in round 1,
we randomly selected half of the polymers
and half of the \acp{AuNP} 
for which we vary 
the plasticity and the temperature, respectively,
and then initialize the machines
from an effectively warm start;
see Figure~S9 in the \ac{SI} as an example.
For each input ($T_{1}$, $T_{2}$),
we obtained the resulting fluxes ($J_{9}$, $J_{10}$)
for each of the 400 \acp{PNNP}
after a round of \ac{MD} simulations.
The 6 plots of ($J_{9}$, $J_{10}$) show similar landscapes 
for the 6 inputs in round 1 because
the initial untrained machines do not discern 
the inputs;
see Figure~S10 in the \ac{SI}.
As a reminder,
the 6 temperature inputs ($T_{1}$, $T_{2}$)
are listed in Table~\ref{tab:input_values} and shown in Figure~S3 in the \ac{SI}.

We now aim to find the best machines among the 400 \acp{PNNP} that can
label the different inputs.
We observe that the distribution of $J_{10}$ has a left shift
with respect to $J_{9}$ in Figure~\ref{fig:round1_fdr}a.
The difference of $|J_9-J_{10}|$ is shown
in Figure~\ref{fig:round1_fdr}b
in directly comparing inputs 1-3 of label 0 
                 with inputs 4-6 of label 1.
We expect that a good \ac{PNNP} machine should lead to a 
broader variation in the
values of the outputs seen in $|J_9-J_{10}|$ 
when comparing outputs for different inputs.
The histograms of $|J_9-J_{10}|$ values for the 6 inputs show similar
distribution patterns in Figure~\ref{fig:round1_fdr}b,
which also means the initial warm start does not discern the inputs.
We calculated the ${\cal F}$ values and plotted the histogram distribution
in Figure~\ref{fig:round1_fdr}c
to quantitatively compare the performance of the \acp{PNNP}.
We found that among the 400 randomly selected warm start settings,
62 of them yield ${\cal F}\ge1$,
which means that these settings can potentially
classify different temperature inputs with label 0 and 1.


\begin{figure}
\includegraphics[clip=true,scale=1,width=0.7\linewidth]{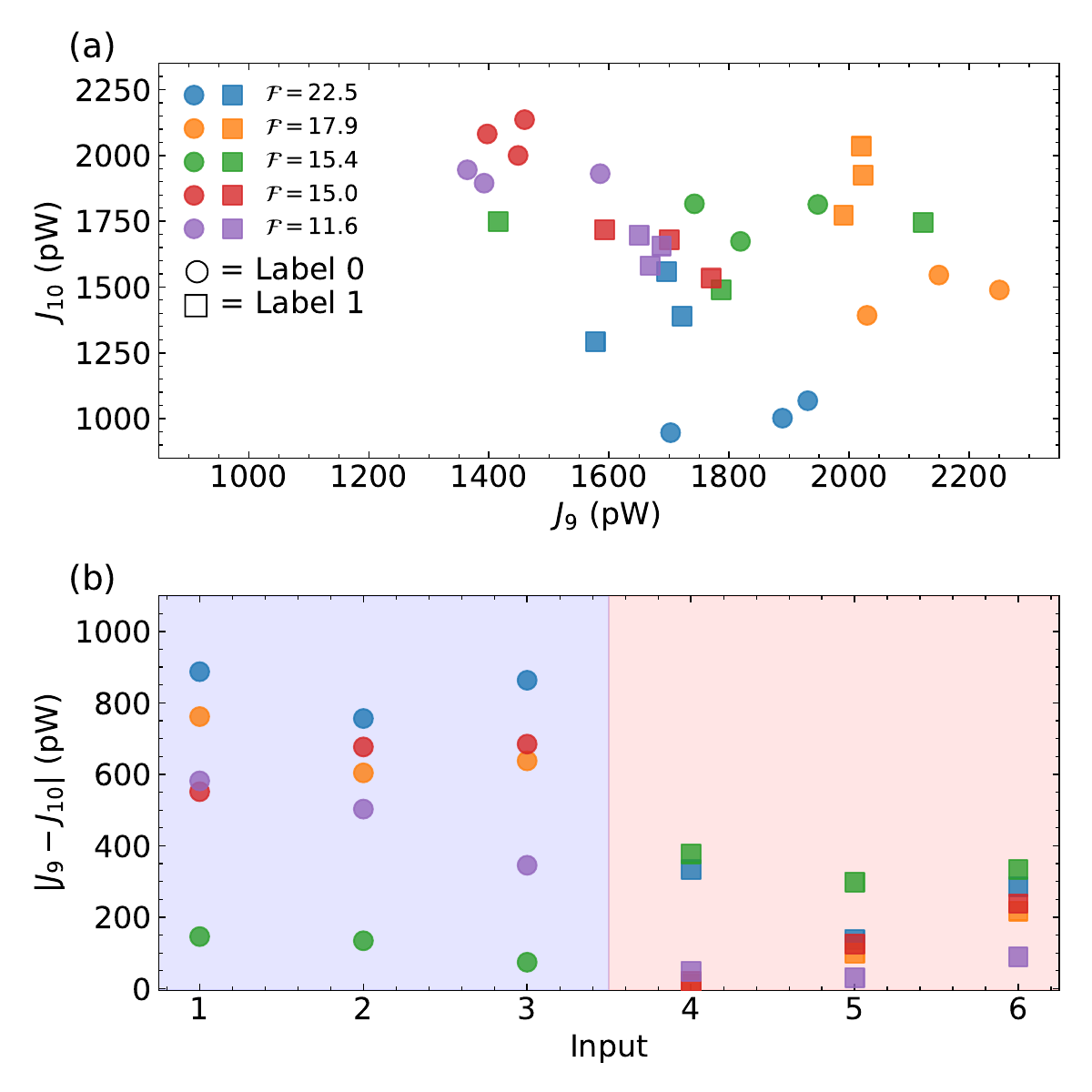}
\caption{
(a) The heat current flow rates at NP9 and NP10, $J_9$ and $J_{10}$,
for the top 5 $\cal F$ values found in round 1 warm start.
(b) The corresponding $|J_9-J_{10}|$ values for input 1-6,
which is sharing the same legend marks and colors as panel (a).
}
\label{fig:round1_top5}
\end{figure}

{\it Selecting the top 5 PNNP machines.}
The top 5 settings ranked by the ${\cal F}$ values in round 1
are the best 5 \ac{PNNP} machines capable of distinguishing inputs 1-3 with label 0
and inputs 4-6 with label 1.
Figure~\ref{fig:round1_top5}a shows that
the 3 pairs of ($J_9$, $J_{10}$) values using inputs 1-3 with label 0 are
separated from those using inputs 4-6 with label 1
for the top 5 settings in round 1.
Specifically,
Figure~\ref{fig:round1_top5}b reveals that
the $|J_9-J_{10}|$ values switch from high to low,
when switching from inputs 1-3 with label 0 to inputs 4-6 with label 1,
for the rank 1, 2, 4, and 5 settings in the top 5 settings.
Meanwhile, for the rank 3 setting,
the $|J_9-J_{10}|$ values switch from low to high,
when inputs switch from label 0 to label 1.
Thus the switching in $|J_9-J_{10}|$ values is
an important but na\"{\i}ve demonstration showing that
the \ac{PNNP} machines can perform classification on different temperature inputs.

\begin{table}[tbp]
\centering
\normalsize 
\caption{Frequency counts of \ac{FDR}, ${\cal F}$, across 5 rounds}
\label{tab:fdr}
\begin{tabular}{lccccc}
\toprule
\textbf{${\cal F}$ range} & \textbf{Round 1} & \textbf{Round 2} & \textbf{Round 3} & \textbf{Round 4} & \textbf{Round 5} \\
\midrule
${\cal F} < 0.1$        & 172 & 160 & 162 & 149 & 140 \\
$0.1 \le {\cal F} < 1$ & 166 & 180 & 172 & 184 & 183 \\
$1 \le{\cal F} < 10$  & 56  & 53  & 65  & 66  & 76  \\
${\cal F} \ge 10$       & 6   & 7   & 1   & 1   & 1   \\
\bottomrule
\end{tabular}
\end{table}

{\it Optimiziing PNNP machines using TuRBO.}
In rounds 2 to 5, we applied the \ac{TuRBO} method to optimize
the temperature settings and polymer plasticities---%
viz.~\{$T_3$, ..., $T_{10}$, $K_1^{1}$, ..., $K_1^{20}$\};
see Figures~S11 to~S26 in the \ac{SI}.
In the \ac{TuRBO} method,
the search for the optimum parameter settings
focuses on dynamically constructed trusted regions of 
the entire parameter space.
It lends itself to strategies, such as the current one,
in which the prior is a good predictor
because it is being successively tuned through 
the previous rounds.
Representative settings of $K_1^3$ and $T_3$ from rounds 2-5 are shown
in Figures~S11, S15, S19, and S23 in the \ac{SI}, respectively.
Although the parameters are not rigorously converging, 
their values do appear to be staying within 
narrowing subdomains.
Although this method is faster than na\"{\i}ve \ac{BO} methods,
it can sometimes miss global optimum values, and some care
was taken to confirm that this was not the case.

In this work,
we aim to demonstrate that the \ac{PNNP} machines can be trained 
by existing \ac{BO} approaches
such as \ac{TuRBO}.
In particular, upon \ac{TuRBO} optimization,
we found that
the distributions of $J_{9}$ and $J_{10}$ values become narrower
as we go through the rounds.
Refer to the figures in the \ac{SI}:
from round 1 warm start in Figure~S10 in the \ac{SI}
through rounds 2 to 5 
in Figures~S12,~S16,~S20, and~S24,
respectively.
The distributions of $|J_9-J_{10}|$ 
also shift from rounds 1 to 5 as the \ac{PNNP} machines evolve.
For example, the peak positions vary across these rounds
in Figures~S13,~S17,~S21, and~S25 in the \ac{SI}.
The \ac{FDR} values also yield more settings with larger ${\cal F}$.
Thus the performance of the \ac{PNNP} machines appear to improve
with subsequent rounds.


Properties of the machines
through the series of \ac{TuRBO} optimizations
upon applying \ac{TuRBO} optimization,
from the round 1 warm start to round 5
are reported in Table~\ref{tab:fdr}.
The number of \ac{PNNP} machines with ${\cal F}\ge 1.0$ 
increases from 62 to 77,
and those with ${\cal F}< 0.1$ decrease from 172 to 140.
These changes in the \ac{FDR}, ${\cal F}$,
provide evidence suggesting that
the desired performance of \ac{PNNP} machines---%
viz.~using heat current to realize information processing
and classify different temperature inputs---%
has been improved.
The top 5 \ac{PNNP} machines in rounds 2 to 5
also exhibit responses switching
between inputs 1-3 (with label 0)
and inputs 4-6 (with label 1)
upon application of different input temperatures.
It yields distinguishable outputs of ($J_{9}$, $J_{10}$),
and leads to a switching in $|J_9-J_{10}|$;
see Figures~S14,~S18,~S22, and~S26 in the \ac{SI}.
The top 5 machine settings used in the \ac{MD} simulations in all 5 rounds
for \{$T_3$, ..., $T_{10}$, $K_1^{1}$, ..., $K_1^{20}$\}
are all available in Table~S1 in the \ac{SI}.

\begin{figure*}
\includegraphics[clip=true,scale=1,width=1\linewidth]{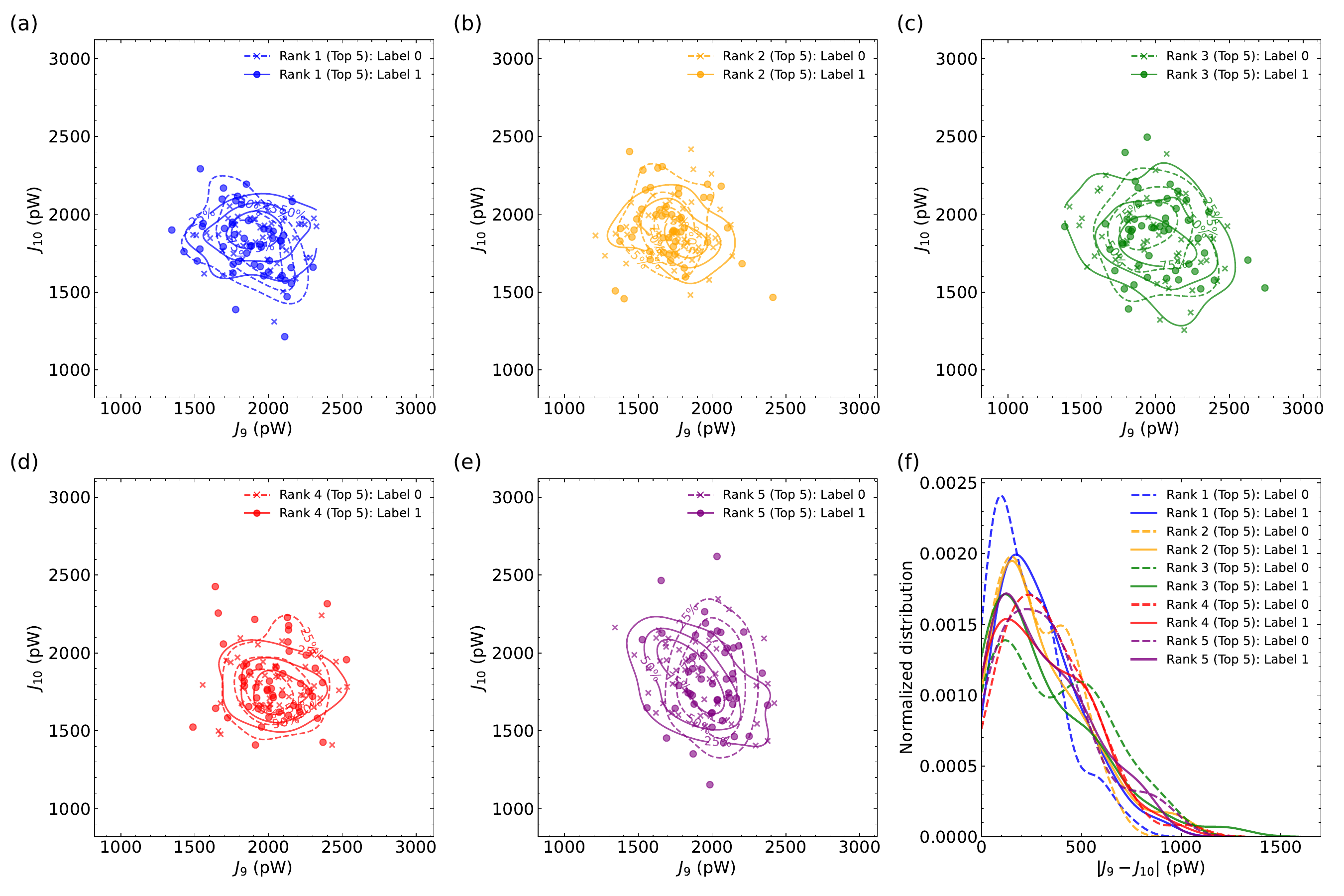}
\caption{
Testing results of the top 5 PNNP machines in round 5 at the end of loop 4
of \ac{MD} simulations in the scheme shown in Figure \ref{fig:scheme}:
(a) to (e)  show the $J_9$ and $J_{10}$ values for rank 1 to 5, respectively.
(f) shows the $|J_9-J_{10}|$ distributions of the two input groups%
---viz. label 0 in dashed lines and label 1 in solid lines---for the top 5 settings.
This testing is using 100 inputs,
where half of them are labelled 0 and the other half are labelled 1.
}
\label{fig:round5_test}
\end{figure*}


{\it Testing top 5 PNNP machines.}
During testing,
we applied 100 pairs of different inputs ($T_1$, $T_2$)
to further evaluate the performance of 
the top 5 \ac{PNNP} machines in each round.
The testing temperature data sets are illustrated in Figure~S27in the \ac{SI},
including the 50 data points with label 0 
and 50 data points with label 1.
For a target \ac{PNNP} machine,
we expect to achieve a significant difference in outputs of ($J_9$, $J_{10}$),
when applying different inputs labeled by 0 or 1.
Panels (a) to (e) in Figure~\ref{fig:round5_test} 
report that for the top 5 settings in round 5,
the plots of ($J_9$, $J_{10}$) have achieved a difference in the landscape
with different inputs, but not a significant one.
Compared to round 1 in Figures~S28 in the \ac{SI},
the plots of  ($J_9$, $J_{10}$) show very similar landscapes 
upon application of inputs labeled by 0 or 1.
From round 1 to round 5,
the top 5 \ac{PNNP} machines exhibit qualitative improvement
in the landscapes of the ($J_9$, $J_{10}$) plots---%
viz. label 0 and label 1 yield different distribution patterns;
see Figures~\ref{fig:round5_test} and ~S28 to~S31 in the \ac{SI}.
Moreover,
Figure~\ref{fig:round5_test}f
also shows that the \ac{PNNP} machines at rank 1-3 of the top 5
contain shoulder peaks at larger $|J_9-J_{10}|$ values in their $|J_9-J_{10}|$ distributions.
This behavior is the key to confirming that
the \ac{PNNP} machines have been improved by the \ac{TuRBO} method after training.
The \ac{PNNP} machines of rank 4-5 of the top 5 also yield 
the $|J_9-J_{10}|$ distributions shown
in Figure~\ref{fig:round5_test}f
that are clearly differentiable.
The $|J_9-J_{10}|$ distributions from round 1 to 4 are available
in panel (f) of Figures~S28 to~S31 in the \ac{SI}.


{\it Conclusions.}
\Acp{PNNP} could one day transform how we approach
computing, \ac{AI} training and data storage.
Here, we use all-atom \ac{MD} simulations
to demonstrate
that \ac{PNNP} can be built to act as \acp{PNN}
capable of performing information process using heat current.
The \ac{PNNP} model emulates a 2-4-2-2 structured \ac{ANN},
which contains 10 nodes (\acp{AuNP}) and 20 edges (polymers).
In the \ac{PNNP} model,
the polymer plasticities and temperatures on \acp{AuNP}
are tuned using high-throughput all-atom \ac{MD} simulations.
By varying the temperatures of the
two 2 input \acp{AuNP},
we performed 5 rounds of simulation to optimize the other 28 parameters,
including a warm start and 4 rounds using the \ac{TuRBO} method.
Our results show that the performance of \ac{PNNP} machines 
improves over the training process,
as we aim to emulate an \ac{ANN}
in labeling nonlinear data sets with data points label 0 and 1.
The \ac{PNNP} machines can process temperature inputs ($T_1$, $T_2$)
and lead to outputs ($J_9$, $J_10$).
We use $|J_9-J_{10}|$ values to measure the difference,
when input data switches from label 0 to 1.
To quantitatively the performance of \ac{PNNP} machines,
the \ac{FDR} values are calculated
for the output $|J_9-J_{10}|$ distributions from label 0 and 1.
The top 5 settings with the largest \ac{FDR} values in these 5 rounds
have been evaluated using 100 testing data as inputs,
which confirms the improvement in performance.

\section*{Methods}
{\it MD simulation model.}
We found that
the fully connected 2-4-2-2 \ac{ANN} model in Figure~\ref{fig:model}a
in capable of classifying 
the nonlinear data set in Figure~S3a in the \ac{SI},
using the tensor flow website \cite{TFplayground}
and Python code.
The 2-4-2-2 \ac{PNNP} all-atom \ac{MD} simulation model in Figure~\ref{fig:model}b
was built to emulate
the \ac{ANN} architecture for information processing using heat currents.
The \ac{PNNP} \ac{MD} simulation models were built in a similar way as
our previous polymer linked \ac{AuNP} models.\cite{hern22l,hern23i}
Namely,
the \ac{PNNP} model contains 10 \acp{AuNP}---each with 4 nm diameters---%
and 20 polyethylene linkers---each with
backbones containing 100 carbons.
For convenience,
the \acp{AuNP} are labeled NP1 to NP10
as indicated in Figure~\ref{fig:model}b.
The simulation is performed in vacuum and the box size is $\sim100\times100\times100$~nm$^3$
with periodic boundary conditions along all 3 dimensions.
The \ac{PNNP} is at the center of the box 
and we place it in a vacuum environment
to suppress energy dissipation.

{\it Simulation force field.}
The interactions of the polymer are represented by
the \ac{OPLS} force field.\cite{jorgensen96, jorgensen01}
The Au-Au interactions in \acp{AuNP} are simulated
by the \ac{LJ} force field of \citet{naik08}.
The polymers are bonded to the \acp{AuNP} through Au-S bonds
described by a Morse potential.\cite{schatz_jang11} 
Other atomistic interactions between organic atoms 
and the Au atoms are represented by
\ac{LJ} potentials applying a weaker interaction energy on Au atoms
according to \citet{schatz_jang11}.
Both \ac{LJ} and Coulombic force cutoff distances
are set to 10~{\AA}~with an additional neighbor list checked at 2.5~{\AA}.
The long-range pairwise interaction energy is not included.
Additional details about the \ac{MD} simulation model can be found 
in our previous papers.\cite{hern22l,hern23i}

{\it Simulation protocol.}
The \ac{LAMMPS} package \cite{plimpton95}
is used to propagate all \ac{MD} simulations.
The simulation timestep is 0.25 fs
which is admittedly short compared to typical timesteps
but necessary here because 
we need to include the vibrational modes from 
the hydrogen atoms.
After the polymers and \acp{AuNP} are connected in the \ac{PNNP} model,
we performed $NVT$ simulations at 300 K to relax 
the structures for more than 5 ns.
The follow-up production run was performed under $NVE$ for 2 ns,
where we applied 10 independent Langevin thermostats on the 10 \acp{AuNP}.
For reference, in Figure~S2 in the \ac{SI},
we first consider a \ac{PNNP} machine in which the temperature of 
both of the \acp{AuNP} in the input
layer, $T_1$ and $T_2$, are set to 300K, and the
temperature gaps
between connected \acp{AuNP} in successive layers
are all set to $\Delta T =50$ K.
The accumulative thermal energy applied to 10 \acp{AuNP} during the simulations
of this \ac{PNNP}
is illustrated in Figure~S2 in the \ac{SI}.

{\it Calculation methods.}
Our target is to maximize the ${\cal F}$ values of $|J_9-J_{10}|$ results
from inputs 1-3 vs inputs 4-6,
which is a simple way to demonstrate that inputs 1-3 and inputs 4-6
can yield different outputs.
Figure~\ref{fig:scheme} shows our scheme for training the \acp{PNNP}.
In all,
we performed 5 rounds of high-throughput \ac{MD} simulations to optimize the 28 parameters.
In each round,
we simulated 400 different settings,
namely 400 \ac{PNNP} machines.
Each machine ran 6 simulations for inputs 1-6
to calculate the corresponding ($J_9$, $J_{10}$) and ${\cal F}$ values,
which requires 2400 \ac{MD} simulations in each round.
In the warm start of round 1, we used 400 randomly selected parameter settings;
see Figure~S9 in the \ac{SI}.
The ($J_9$, $J_{10}$) values were calculated from the \ac{MD} simulations,
which gives $|J_9-J_{10}|$ values.
By grouping the 3 values of $|J_9-J_{10}|$ from inputs 1-3 and 3 from inputs 4-6,
we can calculate the ${\cal F}$ value for each setting.
Using the 400 settings and ${\cal F}$ values from round 1,
we obtained the next 400 settings needed to run round 2
using the \ac{TuRBO} method
maximizing ${\cal F}$.\cite{eriksson19turbo}
Here,
na\"{\i}ve \ac{BO} methods are not appropriate
because we have more than 20 dimensions,\cite{frazier2018BO}
and developing new \ac{BO} methods
is out of scope of the present work.
Instead, we iterate through the procedure illustrated in Figure~\ref{fig:scheme}
to improve the \ac{PNNP} performance
as follows:
Round 2 calculated the another 400 \ac{FDR} values, ${\cal F}$, using the 400 new settings from round 1.
Round 3 used both results from rounds 1 and 2 to generate the next 400 new settings and run simulations.
Round 4 used all results from rounds 1 to 3 to generate the next 400 new settings and run simulations.
Round 5 used all results from rounds 1 to 4 to generate the next 400 new settings and run simulations.
We finished training the \ac{PNNP} machines at round 5 due to computational cost limitation.
Overall,
we performed 12000 high-throughput \ac{MD} simulations in the training process
and 2500 in the testing process.
The total computational cost of this work is about 4.5M CPU SUs on PSC Bridges-2 RM.

\section*{Acknowledgement}
This work has been partially supported by the 
National Science Foundation (NSF) through Grant No.~CHE 2102455.
The computing resources necessary for this work were
performed in part on 
Expanse 
at the San Diego Supercomputing Center
through allocation CTS090079
provided 
by \ac{ACCESS}, which is supported by National Science Foundation (NSF)
grants \#2138259, \#2138286, \#2138307, \#2137603, and \#2138296.
Additional computing resources
were provided by the Advanced Research Computing at Hopkins (ARCH) 
high-performance computing (HPC) facilities supported 
by the NSF MRI Grant (OAC-1920103).

\section*{Supplementary Information}
\label{sec:SI}

The Supplementary Information (SI) includes the following contents:
Figures~S1 to~S8 provide additional information about simulation model and methods.
Figures~S9 to~S26
and Table~S1 provide additional information about  5 rounds of training \acp{PNNP}.
Figures~S27 to~S31 show results of 100 testing inputs.



\FloatBarrier

\newcommand{\doi}[1]{\href{http://dx.doi.org/#1}{\nolinkurl{#1}}}
\bibliography{paper_PNN}

@article{mcculloch43,
  author =        {McCulloch, Warren S and Pitts, Walter},
  journal =       {Bull. Math. Biophys.},
  number =        {4},
  pages =         {115--133},
  publisher =     {Springer},
  title =         {A Logical Calculus of the Ideas Immanent in Nervous
                   Activity},
  volume =        {5},
  year =          {1943},
  doi =           {10.1007/BF02478259},
}

@misc{datamapu23,
  author =        {Datamapu},
  howpublished =  {Towards AI},
  month =         {January 29},
  note =          {Accessed: 20 February 2026},
  title =         {A Brief History of Neural Nets},
  year =          {2023},
  url =           {https://pub.towardsai.net/a-brief-history-of-neural-nets-
                  472107bc2c9c},
}

@article{fleury25,
  author =        {Momeni, Ali and Rahmani, Babak and Scellier, Benjamin and
                   Wright, Logan G. and McMahon, Peter L. and
                   Wanjura, Clara C. and Li, Yuhang and Skalli, Anas and
                   Berloff, Natalia G. and Onodera, Tatsuhiro and
                   Oguz, Ilker and Morichetti, Francesco and
                   del Hougne, Philipp and Le Gallo, Manuel and
                   Sebastian, Abu and Mirhoseini, Azalia and
                   Zhang, Cheng and Markovi{\'c}, Danijela and
                   Brunner, Daniel and Moser, Christophe and
                   Gigan, Sylvain and Marquardt, Florian and
                   Ozcan, Aydogan and Grollier, Julie and Liu, Andrea J. and
                   Psaltis, Demetri and Al{\`u}, Andrea and
                   Fleury, Romain},
  journal =       {Nature},
  number =        {8079},
  pages =         {53--61},
  title =         {Training of Physical Neural Networks},
  volume =        {645},
  year =          {2025},
  doi =           {10.1038/s41586-025-09384-2},
  isbn =          {1476-4687},
}

@incollection{hebb49,
  author =        {Hebb, Donald Olding},
  publisher =     {John Wiley \& Sons},
  title =         {The Organization of Behavior: A Neuropsychological
                   Theory},
  year =          {1949},
}

@article{khan20,
  author =        {Khan, Asifullah and Sohail, Anabia and Zahoora, Umme and
                   Qureshi, Aqsa Saeed},
  journal =       {Artif. Intell. Rev.},
  number =        {8},
  pages =         {5455--5516},
  title =         {A Survey of The Recent Architectures of Deep
                   Convolutional Neural Networks},
  volume =        {53},
  year =          {2020},
  doi =           {10.1007/s10462-020-09825-6},
  isbn =          {1573-7462},
  url =           {https://doi.org/10.1007/s10462-020-09825-6},
}

@article{lecun89,
  author =        {LeCun, Y. and Boser, B. and Denker, J. S. and
                   Henderson, D. and Howard, R. E. and Hubbard, W. and
                   Jackel, L. D.},
  journal =       {Neural Computation},
  number =        {4},
  pages =         {541-551},
  title =         {Backpropagation Applied to Handwritten Zip Code
                   Recognition},
  volume =        {1},
  year =          {1989},
  doi =           {10.1162/neco.1989.1.4.541},
}

@article{alexnet12,
  author =        {Alex Krizhevsky and Ilya Sutskever and
                   Geoffrey E Hinton},
  journal =       {Commun. ACM},
  number =        {6},
  pages =         {84-90},
  title =         {Imagenet Classification with Deep Convolutional
                   Neural Networks},
  volume =        {60},
  year =          {2017},
  doi =           {10.1145/3065386},
}

@inproceedings{brown20gpt3,
  author =        {Brown, Tom and Mann, Benjamin and Ryder, Nick and
                   Subbiah, Melanie and Kaplan, Jared D and
                   Dhariwal, Prafulla and Neelakantan, Arvind and
                   Shyam, Pranav and Sastry, Girish and Askell, Amanda and
                   Agarwal, Sandhini and Herbert-Voss, Ariel and
                   Krueger, Gretchen and Henighan, Tom and Child, Rewon and
                   Ramesh, Aditya and Ziegler, Daniel and Wu, Jeffrey and
                   Winter, Clemens and Hesse, Christopher and Chen, Mark and
                   Sigler, Eric and Litwin, Mateusz and Gray, Scott and
                   Chess, Benjamin and Clark, Jack and
                   Berner, Christopher and McCandlish, Sam and
                   Radford, Alec and Sutskever, Ilya and Amodei, Dario},
  booktitle =     {Advances in Neural Information Processing Systems},
  editor =        {H. Larochelle and M. Ranzato and R. Hadsell and
                   M.F. Balcan and H. Lin},
  pages =         {1877--1901},
  publisher =     {Curran Associates, Inc.},
  title =         {Language Models are Few-Shot Learners},
  volume =        {33},
  year =          {2020},
  url =           {https://proceedings.neurips.cc/paper/2020/file/
                  1457c0d6bfcb4967418bfb8ac142f64a-Paper.pdf},
}

@article{Williams2017,
  author =        {R. Stanley Williams},
  journal =       {Comput. Sci. Eng.},
  number =        {2},
  pages =         {7-13},
  title =         {What's Next? [{T}he End of {M}oore's {L}aw]},
  volume =        {19},
  year =          {2017},
  doi =           {10.1109/MCSE.2017.31},
}

@article{Shalf2020,
  author =        {John Shalf},
  journal =       {Philos. Trans. R. Soc., A},
  number =        {2166},
  pages =         {20190061},
  title =         {The Future of Computing Beyond {M}oore's {L}aw},
  volume =        {378},
  year =          {2020},
  doi =           {10.1098/rsta.2019.0061},
}

@article{Sebastian2020,
  author =        {Abu Sebastian and Manuel Le Gallo and
                   Riduan Khaddam-Aljameh and Evangelos Eleftheriou},
  journal =       {Nat. Nanotechnol.},
  pages =         {529-544},
  title =         {Memory Devices and Applications for In-Memory
                   Computing},
  volume =        {15},
  year =          {2020},
  doi =           {10.1038/s41565-020-0655-z},
}

@article{Pronold2022,
  author =        {J. Pronold and J. Jordan and B.J.N. Wylie and
                   I. Kitayama and M. Diesmann and S. Kunkel},
  journal =       {Parallel Comput.},
  pages =         {102952},
  title =         {Routing Brain Traffic Through the von {N}eumann
                   Bottleneck: Efficient Cache Usage in Spiking Neural
                   Network Simulation Code on General Purpose Computers},
  volume =        {113},
  year =          {2022},
  doi =           {doi.org/10.1016/j.parco.2022.102952},
  issn =          {0167-8191},
}

@article{wetzstein20,
  author =        {Wetzstein, Gordon and Ozcan, Aydogan and
                   Gigan, Sylvain and Fan, Shanhui and Englund, Dirk and
                   Solja{\v c}i{\'c}, Marin and Denz, Cornelia and
                   Miller, David A. B. and Psaltis, Demetri},
  journal =       {Nature},
  number =        {7836},
  pages =         {39--47},
  title =         {Inference in Artificial Intelligence with Deep Optics
                   and Photonics},
  volume =        {588},
  year =          {2020},
  doi =           {10.1038/s41586-020-2973-6},
  isbn =          {1476-4687},
  url =           {https://doi.org/10.1038/s41586-020-2973-6},
}

@techreport{rosenblatt57,
  address =       {Buffalo, NY},
  author =        {Rosenblatt, Frank},
  institution =   {Cornell Aeronautical Laboratory},
  note =          {Project PARA},
  number =        {85-460-1},
  title =         {The Perceptron: A Perceiving and Recognizing
                   Automaton},
  year =          {1957},
}

@article{rosenblatt58,
  author =        {Rosenblatt, Frank},
  journal =       {Psychol. Rev.},
  number =        {6},
  pages =         {386},
  publisher =     {American Psychological Association},
  title =         {The Perceptron: A Probabilistic Model for Information
                   Storage and Organization in The Brain},
  volume =        {65},
  year =          {1958},
  doi =           {10.1037/h0042519},
}

@article{williams2008,
  author =        {Dmitri B. Strukov and Gregory S. Snider and
                   Duncan R. Stewart and R. Stanley Williams},
  journal =       {Nature},
  pages =         {80–83},
  title =         {The Missing Memristor Found},
  volume =        {453},
  year =          {2008},
  doi =           {10.1038/nature06932},
}

@article{Xiao2023,
  author =        {Yongyue Xiao and Bei Jiang and Zihao Zhang and
                   Shanwu Ke and Yaoyao Jin and Xin Wen and Cong Ye and},
  journal =       {Sci. Tech. Adv. Mat.},
  number =        {1},
  pages =         {2162323},
  publisher =     {Taylor \& Francis},
  title =         {A review of memristor: material and structure design,
                   device performance, applications and prospects},
  volume =        {24},
  year =          {2023},
  doi =           {10.1080/14686996.2022.2162323},
}

@article{Weilenmann2024,
  author =        {Christoph Weilenmann and Alexandros N. Ziogas and
                   Tobias Zellweger and others},
  journal =       {Nat. Commun.},
  pages =         {6898},
  publisher =     {Springer Nature},
  title =         {Single neuromorphic memristor closely emulates
                   multiple synaptic mechanisms for energy efficient
                   neural networks},
  volume =        {15},
  year =          {2024},
  doi =           {10.1038/s41467-024-51093-3},
}

@article{mcmahon23a,
  author =        {McMahon, Peter L.},
  journal =       {Nat. Rev. Phys.},
  number =        {12},
  pages =         {717--734},
  title =         {The Physics of Optical Computing},
  volume =        {5},
  year =          {2023},
  doi =           {10.1038/s42254-023-00645-5},
  isbn =          {2522-5820},
  url =           {https://doi.org/10.1038/s42254-023-00645-5},
}

@article{Li2023mem,
  author =        {Dong, Xiaofei and Wei, Wenbin and Sun, Hao and
                   Li, Siyuan and Chen, Jianbiao and Chen, Jiangtao and
                   Zhang, Xuqiang and Zhao, Yun and Li, Yan},
  journal =       {J. Chem. Phys.},
  month =         {05},
  number =        {18},
  pages =         {184702},
  title =         {{Neotype kuramite optoelectronic memristor for
                   bio-synaptic plasticity simulations}},
  volume =        {158},
  year =          {2023},
  doi =           {10.1063/5.0151205},
  issn =          {0021-9606},
}

@article{mcmahon23b,
  author =        {Wang, Tianyu and Sohoni, Mandar M. and
                   Wright, Logan G. and Stein, Martin M. and
                   Ma, Shi-Yuan and Onodera, Tatsuhiro and
                   Anderson, Maxwell G. and McMahon, Peter L.},
  journal =       {Nat. Photon.},
  number =        {5},
  pages =         {408--415},
  title =         {Image Sensing with Multilayer Nonlinear Optical
                   Neural Networks},
  volume =        {17},
  year =          {2023},
  doi =           {10.1038/s41566-023-01170-8},
  isbn =          {1749-4893},
  url =           {https://doi.org/10.1038/s41566-023-01170-8},
}

@article{hern21b,
  author =        {Mark Bathe and Rigoberto Hernandez and
                   Takaki Komiyama and Raghu Machiraju and
                   Sanghamitra Neogi},
  journal =       {ACS Nano},
  number =        {3},
  pages =         {3586–3592},
  title =         {Autonomous Computing Materials},
  volume =        {15},
  year =          {2021},
  doi =           {10.1021/acsnano.0c09556},
}

@article{ElHelou2021,
  author =        {Charles El Helou and Philip R. Buskohl and
                   Christopher E. Tabor and Ryan L. Harne},
  journal =       {Nat. Commun.},
  number =        {1},
  pages =         {1633},
  title =         {Digital Logic gates in Soft, Conductive Mechanical
                   Metamaterials},
  volume =        {12},
  year =          {2021},
  doi =           {10.1038/s41467-021-21920-y},
}

@article{hern21i,
  author =        {Xingfei Wei and Yinong Zhao and Yi Zhuang and
                   Rigoberto Hernandez},
  journal =       {J. Chem. Phys.},
  pages =         {154704},
  title =         {Building Blocks for Autonomous Computing Materials:
                   Dimers, Trimers and Tetramers,},
  volume =        {155},
  year =          {2021},
  doi =           {10.1063/5.0064988},
}

@article{hern23e,
  author =        {Xingfei Wei and Ewa Harazinska and
                   Rigoberto Hernandez},
  journal =       {Phys. Rev. Res.},
  pages =         {L022057},
  title =         {Control of Structure and Dynamics in
                   Polymer-Networked Engineered Nanoparticle Arrays by
                   Electric Fields},
  volume =        {5},
  year =          {2023},
  doi =           {10.1103/PhysRevResearch.5.L022057},
}

@article{hern24k,
  author =        {Yinong Zhao and Xingfei Wei and Rigoberto Hernandez},
  journal =       {J. Phys. Chem. C},
  pages =         {21164-21172},
  title =         {Neuromorphic Computing Primitives Using
                   Polymer-Networked Nanoparticles},
  volume =        {128},
  year =          {2024},
  doi =           {10.1021/acs.jpcc.4c06055},
}

@article{hern25i,
  author =        {Yinong Zhao and Xingfei Wei and Rigoberto Hernandez},
  journal =       {J. Phys. Chem. A},
  number =        {36},
  pages =         {8432-8440},
  title =         {Emergence of Polymer-Networked Nanoparticle
                   Structures as Primitive Neuromorphic Computing
                   States},
  volume =        {129},
  year =          {2025},
  doi =           {10.1021/acs.jpca.5c02941},
}

@article{hern21d,
  author =        {Xingfei Wei and Yinong Zhao and Yi Zhuang and
                   Rigoberto Hernandez},
  journal =       {J. Chem. Phys.},
  pages =         {214702},
  title =         {Engineered Nanoparticle Network Models for Autonomous
                   Computing},
  volume =        {154},
  year =          {2021},
  doi =           {10.1063/5.0048898},
}

@inproceedings{hern24h,
  author =        {Xingfei Wei and Ewa Harazinska and Yinong Zhao and
                   Rigoberto Hernandez},
  booktitle =     {2024 IEEE 24th International Conference on
                   Nanotechnology (NANO)},
  pages =         {409-413},
  title =         {Networked Nanoparticle Arrays for Autonomous
                   Computing: (Invited Paper)},
  year =          {2024},
  doi =           {10.1109/NANO61778.2024.10628712},
}

@article{bli2012,
  author =        {Nianbei Li and Jie Ren and Lei Wang and Gang Zhang and
                   Peter H{\"a}nggi and Baowen Li},
  journal =       {Rev. Mod. Phys.},
  number =        {3},
  pages =         {1045},
  title =         {Colloquium: Phononics: Manipulating Heat Flow with
                   Electronic Analogs and Beyond},
  volume =        {84},
  year =          {2012},
  doi =           {10.1103/RevModPhys.84.1045},
}

@article{bli2020,
  author =        {Kezhao Xiong and Zonghua Liu and Chunhua Zeng and
                   Baowen Li},
  journal =       {Natl. Sci. Rev.},
  number =        {2},
  pages =         {270-277},
  title =         {Thermal-Siphon Phenomenon and Thermal/Electric
                   Conduction in Complex Networks},
  volume =        {7},
  year =          {2020},
  doi =           {10.1093/nsr/nwz128},
}

@article{bli2021,
  author =        {Shuan Wang and Chunhua Zeng and Fengzao Yang and
                   Kezhao Xiong and Baowen Li},
  journal =       {Eur. Phys. J. B},
  pages =         {236},
  title =         {Energy Diffusion of Simple Networks under the
                   Spatiotemporal Thermostats},
  volume =        {94},
  year =          {2021},
  doi =           {10.1140/epjb/s10051-021-00247-z},
}

@article{bli2022,
  author =        {Ya-Fei Ding and Gui-Mei Zhu and Xiang-Ying Shen and
                   Xue Bai and Bao-Wen Li},
  journal =       {Chin. Phys. B},
  month =         {nov},
  number =        {12},
  pages =         {126301},
  title =         {Advances of Phononics in 2012-2022},
  volume =        {31},
  year =          {2022},
  doi =           {10.1088/1674-1056/ac935d},
}

@article{hern22l,
  author =        {Xingfei Wei and Ewa Harazinska and Yinong Zhao and
                   Yi Zhuang and Rigoberto Hernandez},
  journal =       {J. Phys. Chem. C},
  pages =         {18511-18519},
  title =         {Thermal Transport Through Polymer Linked Gold
                   Nanoparticles},
  volume =        {126},
  year =          {2022},
  doi =           {10.1021/acs.jpcc.2c05816},
}

@article{hern23i,
  author =        {Xingfei Wei and Rigoberto Hernandez},
  journal =       {J. Phys. Chem. Lett.},
  pages =         {9834},
  title =         {Heat Transfer Enhancement in Tree-Structured Polymer
                   Linked Gold Nanoparticle Network},
  volume =        {14},
  year =          {2023},
  doi =           {10.1021/acs.jpclett.3c02367},
}

@article{hern24j,
  author =        {Xingfei Wei and Rigoberto Hernandez},
  journal =       {ACS Appl. Mater. Interfaces},
  pages =         {48103–48112},
  title =         {Molecular Electronic Junctions Achieved High Thermal
                   Switch Ratios in Atomistic Simulations},
  volume =        {16},
  year =          {2024},
  doi =           {10.1021/acsami.4c09904},
}

@article{hern25j,
  author =        {Xingfei Wei and Alexander Popov and
                   Rigoberto Hernandez},
  journal =       {J. Phys. Chem. Lett.},
  pages =         {12521–12530},
  title =         {Thermal Switching in a Ferrocenyl Nanojunction Is
                   Observed in All-Atom Simulations},
  volume =        {16},
  year =          {2025},
  doi =           {10.1021/acs.jpclett.5c02789},
}

@inproceedings{eriksson19turbo,
  author =        {Eriksson, David and Pearce, Michael and
                   Gardner, Jacob and Turner, Ryan D and
                   Poloczek, Matthias},
  booktitle =     {Advances in Neural Information Processing Systems 32},
  pages =         {5496--5507},
  title =         {Scalable Global Optimization via Local Bayesian
                   Optimization},
  year =          {2019},
  doi =           {10.48550/arXiv.1910.01739},
  url =           {https://arxiv.org/abs/1910.01739},
}

@article{markussen09,
  author =        {Markussen, Troels and Jauho, Antti-Pekka and
                   Brandbyge, Mads},
  journal =       {Phys. Rev. B},
  month =         {Jan},
  pages =         {035415},
  publisher =     {American Physical Society},
  title =         {Electron and phonon transport in silicon nanowires:
                   Atomistic approach to thermoelectric properties},
  volume =        {79},
  year =          {2009},
  doi =           {10.1103/PhysRevB.79.035415},
  url =           {https://link.aps.org/doi/10.1103/PhysRevB.79.035415},
}

@article{nitzan2023b,
  author =        {Zimbovskaya, Natalya A. and Nitzan, Abraham},
  journal =       {J. Chem. Phys.},
  month =         {06},
  number =        {23},
  pages =         {234903},
  title =         {Phonon Transport Along Long Polymer Chains with
                   Varying Configurations: Effects of Phonon Scattering},
  volume =        {158},
  year =          {2023},
  doi =           {10.1063/5.0155486},
  issn =          {0021-9606},
  url =           {https://doi.org/10.1063/5.0155486},
}

@article{ztian26,
  author =        {Dai, Jinghang and Tian, Zhiting},
  journal =       {MRS Commun.},
  month =         mar,
  publisher =     {Springer Science and Business Media LLC},
  title =         {On the Landauer formula in interfacial thermal
                   transport},
  year =          {2026},
  doi =           {10.1557/s43579-026-00941-y},
  issn =          {2159-6867},
  url =           {http://dx.doi.org/10.1557/s43579-026-00941-y},
}

@article{jliu2012,
  author =        {Jun Liu and Ronggui Yang},
  journal =       {Phys. Rev. B},
  number =        {10},
  pages =         {104307},
  title =         {Length-Dependent Thermal Conductivity of Single
                   Extended Polymer Chains},
  volume =        {86},
  year =          {2012},
  doi =           {10.1103/PhysRevB.86.104307},
}

@article{tluo2012,
  author =        {Teng Zhang and Tengfei Luo},
  journal =       {J. Appl. Phys.},
  number =        {9},
  pages =         {094304},
  title =         {Morphology-Influenced Thermal Conductivity of
                   Polyethylene Single Chains and Crystalline Fibers},
  volume =        {112},
  year =          {2012},
  doi =           {10.1063/1.4759293},
}

@article{nitzan2020,
  author =        {Mohammadhasan Dinpajooh and Abraham Nitzan},
  journal =       {J. Chem. Phys.},
  number =        {16},
  pages =         {164903},
  title =         {Heat Conduction in Polymer Chains with Controlled
                   End-to-End Distance},
  volume =        {153},
  year =          {2020},
  doi =           {10.1063/5.0023085},
}

@article{nitzan2022,
  author =        {Mohammadhasan Dinpajooh and Abraham Nitzan},
  journal =       {J. Chem. Phys.},
  number =        {14},
  pages =         {144901},
  title =         {Heat Conduction in Polymer Chains: Effect of
                   Substrate on the Thermal Conductance},
  volume =        {156},
  year =          {2022},
  doi =           {10.1063/5.0087163},
}

@misc{TFplayground,
  author =        {Smilkov, Daniel and Carter, Shan and
                   Google Brain Team},
  howpublished =  {\url{https://playground.tensorflow.org}},
  note =          {Accessed: 2025-03-04},
  title =         {A Neural Network Playground},
  year =          {2016},
}

@article{jorgensen96,
  author =        {William L. Jorgensen and David S. Maxwell and
                   Julian Tirado-Rives},
  journal =       {J. Am. Chem. Soc.},
  number =        {45},
  pages =         {11225-11236},
  title =         {Development and Testing of the {OPLS} All-Atom Force
                   Field on Conformational Energetics and Properties of
                   Organic Liquids},
  volume =        {118},
  year =          {1996},
  doi =           {10.1021/ja9621760},
}

@article{jorgensen01,
  author =        {George A. Kaminski and Richard A. Friesner and
                   Julian Tirado-Rives and William L. Jorgensen},
  journal =       {J. Phys. Chem. B},
  number =        {28},
  pages =         {6474-6487},
  title =         {Evaluation and Reparameterization of the {OPLS-AA}
                   Force Field for Proteins via Comparison with Accurate
                   Quantum Chemical Calculations on Peptides},
  volume =        {105},
  year =          {2001},
  doi =           {10.1021/jp003919d},
}

@article{naik08,
  author =        {Hendrik Heinz and R. A. Vaia and Barry L. Farmer and
                   R. R. Naik},
  journal =       {J. Phys. Chem. C},
  pages =         {17281-17290},
  title =         {Accurate Simulation of Surfaces and Interfaces of
                   Face-Centered Cubic Metals Using 12-6 and 9-6
                   {Lennard-Jones} Potentials},
  volume =        {112},
  year =          {2008},
  doi =           {10.1021/jp801931d},
}

@article{schatz_jang11,
  author =        {Yoonho Ahn and Joyanta K. Saha and George C. Schatz and
                   Joonkyung Jang},
  journal =       {J. Phys. Chem. C},
  pages =         {10668-10674},
  title =         {Molecular Dynamics Study of the Formation of a
                   Self-Assembled Monolayer on Gold},
  volume =        {115},
  year =          {2011},
  doi =           {10.1021/jp200447k},
}

@article{plimpton95,
  author =        {Steven J. Plimpton},
  journal =       {J. Comput. Phys.},
  number =        {1},
  pages =         {1--19},
  title =         {Fast Parallel Algorithms for Short-Range Molecular
                   Dynamics},
  volume =        {117},
  year =          {1995},
  doi =           {10.1006/jcph.1995.1039},
}

@misc{frazier2018BO,
  author =        {Peter I. Frazier},
  title =         {A Tutorial on Bayesian Optimization},
  year =          {2018},
  url =           {https://arxiv.org/abs/1807.02811},
}
\end{document}